\documentstyle[11pt]{article}

\newlength{\oline}
\newlength{\tline}
\newcommand{\ol}{\vspace{\oline}}

\begin{document}
\begin{flushright}
26 September 1994\\
IC/94/285
\end{flushright}
\begin{center}
{\large \bf ELECTROMAGNETIC-GRAVITATIONAL\\
 CONVERSION CROSS SECTIONS\\
IN  EXTERNAL ELECTROMAGNETIC FIELDS\\}
\ol
{\bf Hoang Ngoc Long\\}
Institute of
Theoretical Physics\\
National Center for Natural Science and Technology\\
 P.O.Box 429, Boho,
Hanoi 10000, Vietnam\\ \ol
{\bf Dang Van Soa\\}
Department of Physics, Hanoi University of Mining and Geology,\\
Dong Ngac, Tu Liem, Hanoi, Vietnam.\\
\ol
and\\
\ol
{\bf Tuan A. Tran\\}
Graduate College in Physics\\
National Center for Natural Science and Technology\\
P.O.Box 429, Boho, Hanoi 10000, Vietnam\\
and\\
International Centre for Theoretical Physics, Trieste, Italy\footnote{
Present address and e-mail: tatran@ictp.trieste.it}\\
\vspace{0.5cm}
{\bf Abstract\\}
\end{center}
\hspace*{1cm}The classical processes: the conversion of photons into
gravitons in the static electromagnetic fields are considered by using
Feynman perturbation techniques. The
differential cross sections are presented for the conversion in the
electric field of the flat condesor and the magnetic field of the selenoid.
A numerical evaluation shows that the cross sections may
have the observable value
in the present technical scenario.\\
\begin {center} PACS numbers : 04.30+x, 04.40.+c, 04.80.+z \end{center}
\newpage
1. {\bf Introduction}\\
\hspace*{1cm}The fundamental consequence of the relativistic theory of
gravitation is the existence of gravitational waves(GWs). Einstein
was the first
to investigate them~\cite{aeistein}. For a long time ( up to the early sixties
)
gravitational radiation has been thought of as a phenomennon of great
theoretical interest, but of no relevance to the real world. However, the
Dyson's remark~\cite{fjdyson} about the GW " energy flux " from
star $F_{gw}\sim 3.629\times 10^{59}$ erg/sec gave an extra stimulus to
consider GWs as real entities. Because of technical
restrictions one mainly pays attention to the outer- space sources of
GWs such as supernovas, binary stars, etc~\cite{dgblair} . However
these experiments are partly passive, because they depend on objective
conditions. \\
\hspace*{1cm}The study on the interaction between electromagnetic (EM) and
gravitational fields is a significant work of research on gravitational
radiation. At the present technical level in the laboratory, it has
been proved by means of mechanics that it is as yet difficult to
generate GWs which are strong enough to be detected by current
detectors. Therefore physicists have transfered their interests to
the strong EM field and take it as one of the possible sources in the
laboratory.\\
\hspace*{1cm}At present, the best hope is to take the interaction
between GW and the EM field as a basis for new methods for detecting
GWs. These new possible methods would be especially suited to GW with
frequencies so high that they cannot be detected by mechanical
antenna, which are limited by the properties of the antenna material.
The EM detection of the gravitons has
been considered by many authors~\cite{wkdelogi}-\cite{hnlong}.
In Ref.~\cite{galupatov} many properties of the EM field
which is generated by a charged condensor in the field of
a GW were considered by using purely classical
method. The results have a methodical meaning only since the size
of the condensor was assumed to be infinite, however, some of them
coincide with ours. In Ref.~\cite{hnlong} we have considered the conversion
of gravitons into photons in the periodic external EM
field by using Feynman diagram techniques. Cross sections have been
calculated in the quasi-static limit. \\
\hspace*{1cm}Here we lay out the theoretical
principles of the most hopeful effect in detection of the gravitons
in eathly conditions : it is the generation of gravitons by
photons in an external EM field. The main advantages of this
effect are following : first, it is the first order of the
perturbation theory, second, since
the EM field is classical theory we can increase
cross sections as much as possible by increasing the intensity of the field or
the volume containing the field.\\
\hspace*{1cm}The paper is organized as follows: some notations and
general matrix element are given in section 2. We consider the
process in static electric field in section 3 and  the process in
magnetic field in section 4. Finally in section 5 we present our conclusions.\\
2. {\bf Photoproduction of gravitons}\\
\hspace*{1cm}Let us consider the process in which the initial state has
the photon $\gamma$ with momentum q and the external electromagnetic field
($EM_{ex}$) and the final state has the graviton
$g$ with momentum $p$ and the above mentioned electromagnetic field :
\begin{equation}
\gamma + EM_{ex}  \longrightarrow g + EM_{ex}.
\label{pemg}
\end{equation}
We shall work with the linear approximation and use the quantization of
Gupta~\cite{sngupta}.\\
\hspace*{1cm}From the interaction Lagrangian for the gravitational field
with the EM field in the linear approximation, the
interaction Lagrangian corresponding to this process has the following
form~\cite{wkdelogi}-\cite{galupatov}

\[L_{int}( A, F_{class},g )=\frac{\kappa}{2}\eta_{\nu\beta}
h_{\mu\alpha}F^{\mu\nu}F^{\alpha\beta}_{class}\]
where $h_{\mu\alpha}$ represents gravitational field, $F_{\mu\nu}$ -
EM strength tensor, $F^{\alpha\beta}_{class}$ - external
EM strength tensor, $\eta_{\nu\beta}$ = $\mbox{diag}( 1, - 1, - 1,
- 1 )$, $\kappa=\sqrt{16\pi G}$.\\
\hspace*{1cm}Using the Feynman rules we get the following expression for
the matrix element~\cite{wkdelogi}-\cite{galupatov}
\begin{equation}
\langle p| M_{ex} |q\rangle=
\frac{\kappa}{4(2\pi)^{2}q_{0}}\varepsilon
^{\lambda}(\vec{q},\sigma)\varepsilon^{\mu\alpha}(\vec{p},
\tau)(\eta_{\lambda\beta}q_{\mu}-q_{\beta}\eta_{\mu\lambda})
\int_{V} e^{i(\vec{q}-\vec{p})\vec{r}}F_{\alpha class}
^{\beta}d\vec{r}
\label{pmexq}
\end{equation}
where $\varepsilon^{\lambda}(\vec{q},\sigma)$ and $\varepsilon^{\mu\alpha}
(\vec{p},\tau)$ represent the polarization tensor of the photon and the
graviton, respectively.\\
\hspace*{1cm}Expression~(\ref{pmexq}) is valid for an arbitrary external
EM field. In the following we shall use it for two cases,
namely the generation in the electric field of a flat condensor and in the
static magnetic field of the selenoid. Here we use the following
notations: $q \equiv |
\vec{q} |,  p \equiv |\vec{p}|$ and $\theta$ is the angle between
$\vec{p}$ and $\vec{q}$. Note that since both photon and graviton are
massless, the energy conservation gives us $p = q$.\\
3. {\bf Photoproduction in electric field}\\
\hspace*{1cm}Let us  consider the generation of gravitons in the
homogeneous electric field of a flat condensor of size $a{\times} b{\times}
c$. We shall use the coordinate system with the x axis parallel to the
direction of the field, i.e.,
\[F^{10}=-F^{01} = E.\]
Then the matrix element is given by
\begin{equation}
\langle p| M_{e} |q\rangle = -\frac{\kappa}{4(2\pi)^{2}}\varepsilon^{i}
(\vec{q},\sigma)\varepsilon_{i1}(\vec{p},\tau)E(\vec{q}-\vec{p}),
i = 1, 2, 3
\label{pmeq}
\end{equation}
where
\[E(\vec{q}-\vec{p}) = \int_{V} e^{i(\vec{q}-\vec{p})\vec{r}}E(\vec{r})
d\vec{r}.\]
For a homogeneous field of intensity $E$ we have
\begin{equation}
E(\vec{q}-\vec{p}) = 8E\frac{\sin[\frac{1}{2}a(q_{x}-p_{x})]\sin[\frac{1}{2}b
(q_{y}-p_{y})]\sin[\frac{1}{2}c(q_{z}-q_{z})]}{(q_{x}-p_{x})(q_{y}-p_{y})
(q_{z}-p_{z})}.
\label{eqmp}
\end{equation}
In order to distinguish the processes in an electric field with those in a
magnetic field, a subscript $e$ has been supplemented to $M$ :  $M_{e}$ .\\
\hspace*{1cm}Squaring the matrix element we meet the expression of summing
up over the polarizations of photons and gravitons. Using the following
well- known formulas~\cite{gpapini}
\[t^{ij}(k)\equiv \sum_{\sigma}\varepsilon^{i}(\vec{k},\sigma)\varepsilon^{
j}(\vec{k},\sigma) = \delta^{i}_{j}-\frac{k^{i}k^{j}}{k^{2}}\]
\vspace*{0.2cm}
\[t^{ij,mn}(k)\equiv \sum_{\sigma}\varepsilon^{ij}(\vec{k},\sigma)
\varepsilon^{mn}(\vec{k},\sigma) = - t^{ij}(k)t^{mn}(k) + t^{im}(k)t^{jn}
(k) + t^{in}(k)t^{jm}(k)\]
\begin{flushright}
$i, j, m, n = 1, 2, 3.$
\end{flushright}
we find easily
\begin{equation}
\sum_{\sigma\tau}\varepsilon^{i}(\vec{q},\sigma)\varepsilon^{j}(\vec{q},
\sigma)\varepsilon_{i1}(\vec{p},\tau)\varepsilon_{j1}(\vec{p},\tau) =
\left( 1 - \frac{p^{2}_{x}}{p^{2}}\right)( 1 + \cos^{2}\theta )
\label{4epsilon}
\end{equation}
\hspace*{1cm}From Eq.~(\ref{pmeq}) and Eq.~(\ref{4epsilon}) the differential
cross section for this process is given
\begin{equation}
\frac{d^{e}\sigma(\gamma \rightarrow g)}{d\Omega}=\frac{\kappa^{2}}{32(2
\pi)^{2}}|E(\vec{q}-\vec{p})|^{2}\left( 1 - \frac{p^{2}_{x}}{p^{2}}\right)
q^{2}(1+\cos^{2}\theta)
\label{desigma}
\end{equation}
Finally substituting Eq.~(\ref{eqmp}) into Eq.~(\ref{desigma}) we find the
differential cross
section of the generation of gravitons in the electric field of a
flat condensor of size $ a\times b\times c$
\begin{eqnarray}
\frac{d^{e}\sigma(\gamma \rightarrow g)}
{d\Omega}&=&\frac{2\kappa^{2} E^{2}}{(2\pi)^{2}}
\left[\frac{\sin(\frac{1}{2}a
(q_{x}-p_{x}))\sin(\frac{1}{2}b(q_{y}-p_{y}))
\sin(\frac{1}{2}c(q_{z}-q_{z}))}
{(q_{x}-p_{x})(q_{y}-p_{y})(q_{z}-p_{z})}\right]^{2} \nonumber \\
         &  &\times q^{2}( 1 + \cos^{2}\theta )\left( 1 -
\frac{p^{2}_{x}}{p^{2}}\right).
\label{desigmap}
\end{eqnarray}
\hspace*{1cm}Let us consider the following cases:\\
\hspace*{0.5cm}a) The momentum of photon is parallel to the z axis, i.e.,
$q^{\mu} =( q, 0, 0, q )$. We have then
\begin{equation}
p_{x}=q\sin\theta \cos\varphi,\ p_{y}=q\sin\theta\sin\varphi ,\
p_{z}=q\cos\theta.
\label{zaxis}
\end{equation}
where $\varphi$ is the angle between the x axis and the projection of
$\vec{p}$ on the xy plane.\\
\hspace*{1cm}Substitution of Eq.~(\ref{zaxis}) into Eq.~(\ref{desigmap})
we get
\begin{eqnarray}
\frac{d^{e}\sigma(\gamma \rightarrow g)}
{d\Omega}&=&\frac{2\kappa^{2}E^{2}}{(2\pi)^{2}q^{4}}
\left[\sin\frac{aq\sin \theta \cos \varphi}{2}
\sin\frac{bq\sin \theta \sin \varphi}{2}
\sin\frac{cq( 1- \cos\theta )}{2}\right]^{2} \nonumber \\
         & &\times [\sin^{2} \theta \sin \varphi \cos \varphi (1-\cos
\theta)]^{-2}( 1 + \cos^{2} \theta )( 1 - \sin^{2} \theta \cos^{2} \varphi).
\label{desigmaz}
\end{eqnarray}
{}From Eq.~(\ref{desigmaz}) we have
\begin{equation}
\frac{d^{e}\sigma(\gamma \rightarrow g)}{d\Omega}=\frac{\kappa^{2}E^{2}}
{16(2\pi)^{2}}V^{2}q^{2},\hspace*{1cm}V \equiv a.b.c
\label{desaeo}
\end{equation}
for $\theta\approx 0$  ( $\theta\ll\frac{2}{aq} , \frac{2}{bq}, \frac{2}{
\sqrt{cq}} )$    and
\begin{equation}
\frac{d^{e}\sigma(\gamma \rightarrow g)}{d\Omega} = \frac{\kappa^{2}E^{2}
a^{2}}{2(2\pi)^{2}q^{2}}\sin^{2}\frac{bq}{2}\sin^{2}\frac{cq}{2}
\label{desaepi}
\end{equation}
for $\theta$=$\frac{\pi}{2}$, $\varphi$=$\frac{\pi}{2}$ and
\begin{equation}
\frac{d^{e}\sigma(\gamma \rightarrow g)}{d\Omega}=0
\label{desaphieo}
\end{equation}
for $\theta$=$\frac{\pi}{2}$, $\varphi$=0.

\hspace*{1cm}Expressions~(\ref{desaeo})-(\ref{desaphieo}) show that the
probabilities of the
gravitons generation are largest in the direction of photon
motion. Formula~(\ref{desaeo}) shows that the differential cross section
depends quadratically on the square of the intensity of the electric
field $E$, the volume $V$ of the condensor and the photon momentum $q$.
Expression~(\ref{desaepi}) shows that in order to get $\sigma\approx 10^{-30}
\mbox{cm}^{2}$ one has to use
an electric field of intensity $E\approx 10^{10}$/a$\lambda$, while
expression~(\ref{desaeo})
shows that in order to get the same cross section we need an
electric field of intensity $E\approx 10^{9}\lambda$/V  only.
When $q\rightarrow 0$ the right hand  side of Eq.~(\ref{desaepi})
is proportional to $q^{2}$.\\
\hspace*{0.5cm}b) The momentum of photon is parallel to the x axis, i.e.,
$q^{\mu} =( q, q, 0, 0 )$. \\
\hspace*{1cm}In the spherical coordinate system the components of
$\vec{p}$ are given as follows
\begin{equation}
p_{x}=q\cos\theta ,\ p_{y}=q\sin\theta \cos\varphi',\
p_{z}=q\sin\theta\sin\varphi'.
\label{xaxis}
\end{equation}
where $\varphi'$ is the angle between the y axis and the projection of
$\vec{p}$  on the yz plane. Substituting Eq.~(\ref{xaxis})
into Eq.~(\ref{desigmap}) we get
\begin{eqnarray}
\frac{d^{e}\sigma (\gamma \rightarrow g)}
{d\Omega'}&=&\frac{2\kappa^{2}E^{2}}{(2\pi)^{2}q^{4}}
\left[\sin\frac{aq(1-\cos\theta)}{2}
\sin\frac{bq\sin\theta \cos\varphi'}{2}
\sin\frac{cq\sin\theta \sin\varphi'}{2}\right]^{2} \nonumber \\
          & &\times [\sin^{2}\theta \sin\varphi' \cos\varphi'
(1-\cos\theta)]^{-2}( 1 - \cos^{4}\theta ).
\label{dopx}
\end{eqnarray}
{}From Eq.~(\ref{dopx}) we have
\[\frac{d^{e}\sigma(\gamma \rightarrow g)}{d\Omega'} =  0\]
for $\theta\approx 0$  ( $\theta\ll\frac{2}{aq},
\frac{2}{bq}, \frac{2}{\sqrt{cq}} )$    and
\begin{equation}
\frac{d^{e}\sigma(\gamma \rightarrow g)}{d\Omega'} = \frac{\kappa^{2}
E^{2}b^{2}}{2(2\pi)^{2}q^{2}}\sin^{2}\frac{aq}{2}\sin^{2}\frac{cq}{2}
\label{dopphiepi}
\end{equation}
for $\theta$=$\frac{\pi}{2}$, $\varphi'$=$\frac{\pi}{2}$ and
\begin{equation}
\frac{d^{e}\sigma(\gamma \rightarrow g)}{d\Omega'}=\frac{\kappa^{2}E^{2}
c^{2}}{2(2\pi)^{2}q^{2}}\sin^{2}\frac{aq}{2}\sin^{2}\frac{bq}{2}
\label{dopphieo}
\end{equation}
for $\theta$=$\frac{\pi}{2}$, $\varphi'$=0.\\
{}From Eqs.~(\ref{dopphiepi}) and~(\ref{dopphieo}) we see that if $b = c$
the probabilities of the
graviton generation are the same in the y- and z- directions, and
if $a = b = c$ the r.h.sides of Eqs.~(\ref{desaepi}),~(\ref{dopphiepi}),
and~(\ref{dopphieo}) are the same. The differential cross-sections in
Eq.~(\ref{dopphiepi}) and Eq.~(\ref{dopphieo}) also depend on
quadratically on the square of the intensity the same as in
Eqs.~(\ref{desaeo}) and~(\ref{desaepi}).\\
4. {\bf Photoproduction in
magnetic field }\\ \hspace*{1cm}Assumming that the direction
of the magnetic field is parallel to the z axis, i.e.,\\
\begin{equation}
F^{12}=-F^{21}=B.
\end{equation}
For the above mentioned process we get the matrix element:
\begin{eqnarray}
\langle p| M^m_{g\gamma} |q\rangle&=&
\frac{\kappa}{4(2\pi)^{2}q_0 }B(\vec{p}-\vec{q})\varepsilon
^i(\vec{q},\sigma)[
\varepsilon^{j1}(\vec{p},\tau)(\eta_{i2}q_j-q_2\eta_{ji})
-\nonumber\\
                                  & &\varepsilon^{j2}(\vec{p},
\tau)(\eta_{i1}q_j-\eta_{ji}q_1)]
\label{pmmq}
\end{eqnarray}
where
\begin{equation}
B(\vec{q}-\vec{p})=\int_{V} e^{i(\vec{q}-\vec{p})\vec{r}}B d\vec{r}.
\label{bqmp}
\end{equation}
Now we calculate integral~(\ref{bqmp}). Suppose that the magnetic field is
homogeneous in the selenoid with the radius $R$ and the lenght $h$.
In the cylindrical coordinates, it becomes\\
\begin{equation}
B(\vec{q}-\vec{p})=B\int_{0}^{R}\varrho d\varrho\int_{0}^{2\pi}\mbox{exp}\{i
[(q_x-p_x)\cos\varphi+(q_y-p_y)\sin\varphi]\}\int_{-h/2}^{h/2}\mbox{exp}[i
(q_z-p_z)z] dz
\label{bqmp1}
\end{equation}

After some manupulations Eq.~(\ref{bqmp1}) can be written as follows
\begin{equation}
B(\vec{q}-\vec{p})=\frac{4\pi BR}{\sqrt{n_x^2+n_y^2}(p_z - q_z)}J_1\left(R
\sqrt{n_x^2+n_y^2}\right)\sin\left(\frac{h(p_z-q_z)}{2}\right).
\label{bqmp2}
\end{equation}
where $n_x\equiv p_x-q_x ,  n_y\equiv p_y-q_y $
and $J_{1}$ is the one-order spherical Bessel function~\cite{fwjolwer}.\\
Substituting~(\ref{bqmp2}) into~(\ref{pmmq}) we find after cumbersome
calculations\\
\begin{eqnarray}
\frac{d\sigma^m(\gamma\rightarrow g)}
{d\Omega}&=&\frac{\kappa^{2}R^{2}B^{2}\sin^{2}\left[\frac{(q_z-p_z)h}{2}\right]
J_{1}^{2}\left(R\sqrt{(q_x-p_x)^{2}+(q_y-p_y)^{2}}\right)}{16(q_z-p_z)^{2}
\left[(q_x-p_x)^{2}+(q_y-p_y)^{2}\right]}\nonumber\\
&\times&
\left((2q^{2}-p_x^{2}-p_y^{2})\sin^{2}\theta+4(p_xq_x+p_yq_y)\cos\theta
-\frac{4}{q^{2}}q_xp_xq_yp_y\right. \nonumber\\
&-&\left. 2(q_x^{2}+q_y^{2})+\frac{2}{q^{2}}(q_x^{2}p_y^{2}
+q_y^{2}p_x^{2})\right).
\label{dmso}
\end{eqnarray}
\hspace*{1cm}When the momentum of the graviton is parallel to the z
axis (the direction of the magnetic field), the differential cross
section vanishes for $\theta=0$ and\\
\begin{equation}
\frac{d\sigma^m(\gamma\rightarrow g)}{d\Omega}=\frac{\kappa^{2}R^{2}
B^{2}}{16q^{2}}
\sin^{2}\left(\frac{qh}{2}\right)J_{1}^{2}(Rq)
\label{dmsoaepi}
\end{equation}
for $\theta = \frac{\pi}{2}$.\\
\hspace*{1cm}Now we consider the case in which the momentum of the
graviton is parallel to the x axis, i.e., $q^{\mu}=(q,q,0,0)$.
Substituting~(\ref{xaxis}) into~(\ref{dmso}) and note that
\[\lim_{p\to q}\frac{J_{1}\left(R(p-q)\right)}{p-q}=\frac{R}{2}\] we
find\\
\begin{equation}
\frac{d\sigma^m(\gamma\rightarrow g)}{d\Omega'}=\frac{\kappa^{2}V^{2}B^{2}
p^{2}}{128\pi^{2}}
\label{dopaeo}
\end{equation}
for $\theta=0$ and\\
\begin{equation}
\frac{d\sigma^m(\gamma\rightarrow g)}{d\Omega'}=\frac{\kappa^{2}h^{2}R^{2}
B^{2}}{128}J^2_1(Rp\sqrt{2})
\label{dopphieop}
\end{equation}
for $\theta=\frac{\pi}{2}$, $\varphi'=0$ and\\
\begin{equation}
\frac{d\sigma^m(\gamma\rightarrow g)}{d\Omega'}=0
\label{dopphiepip}
\end{equation}
for $\theta=\frac{\pi}{2}$, $\varphi'=\frac{\pi}{2}$.\\
\hspace{1cm}From Eq.~(\ref{dopaeo}) we see that the differential cross
section in the direction of graviton motion depends quadratically on
the magnitude $B$, the volume $V$ of the selenoid and the graviton momentum
$p$.\\
\hspace*{1cm}From Eq.~(\ref{dopphieop}) it follows that the differential
cross section vanishes when $p_{n}=\frac{\mu_{n}}{R\sqrt{2}}$ with
$n=0,\pm1\pm2$...and has its largest value
\begin{equation}
\frac{d\sigma^m(g\to\gamma)}{d\Omega'}\approx3.2\times10^{-50}h^{2}R^{2}
B^{2}J_{1}^{2}(\mu'_{n})
\label{doplv}
\end{equation}
for $p_{n}=\frac{\mu'_{n}}{R\sqrt{2}}$. Where $\mu_{n}$ and $\mu'_{n}$
are the roots of $J_{1}(\mu_{n})=0$ and $J'_{1}(\mu'_{n})=0$.\\
5. {\bf Discussion}\\
\hspace*{1cm}The following consequences may be obtained from our results\\
\hspace*{0.5cm}1 - It is the best for experiments when the momentum of
photons is perpendicular to the EM field and in this case, the
conversion cross sections are largest in the
direction of the photon motion.\\
\hspace*{0.5cm}2 - In C.G.S units, formulas~(\ref{desaeo}) and~(\ref{desaepi})
have the
forms, respectively
\begin{equation}
\frac{d\sigma_{\parallel}(\gamma \rightarrow g)}
{d\Omega}\approx 1.32.10^{-49}\frac{V^{2}E^{2}}{\lambda^{2}}
\label{dspara}
\end{equation}
\vspace*{0.2cm}
\begin{equation}
\frac{d\sigma_{\perp}(\gamma \rightarrow g)}
{d\Omega}\approx 1.32.10^{-51}a^{2}E^{2}\lambda^{2}
\sin^{2}\frac{b\pi}{\lambda}\sin^{2}\frac{c\pi}{\lambda}
\label{dsper}
\end{equation}
Note that when $\lambda\gg b, c$~(\ref{dspara}) becomes
\[\frac{d\sigma_{\perp}(\gamma \rightarrow g)}
{d\Omega}\approx 1.32.10^{-49}\frac{V^{2}E^{2}}{\lambda^{2}}\]
Suppose that the size of the condensor is, $1m\times 1m\times 1m$, the
intensity of the electric field  $E = 100\frac{kV}{m}$ and the photon
length $\lambda$ = $10^{-5}$ cm, the cross section given by~(\ref{dspara})
is
$\frac{d\sigma_{\parallel}(\gamma{\rightarrow}g)}{d\Omega}\approx 10^{-16}
\mbox{cm}^{2}$, while $\frac{d\sigma_{\perp}(\gamma{\rightarrow}g)}
{d\Omega}\approx 10^{-46} \mbox{cm}^{2}$. The situation is analogous in the
case
of the magnetic field.\\
In C. G. S. units Eq.~(\ref{dopaeo}) becomes
\begin{equation}
\frac{d\sigma(g\to\gamma)}{d\Omega'}\approx 1.3\times
10^{-49}\frac{V^{2}B^{2}}{\lambda^{2}}
\end{equation}
where $\lambda$ is the wavelength of graviton and the cross section
gives
\begin{equation}
\frac{d\sigma(g\to\gamma)}{d\Omega'}\approx1.3\times10^{-15} \mbox{cm}^{2}
\end{equation}
for $V=10^{6} \mbox{cm}^{3}$ ,$B=10^{6}
\mbox{cm}^{-1/2}\mbox{g}^{1/2}\mbox{s}^{-1}$ , $\lambda=10^{-5}$ cm \\
\hspace*{0.5cm}3 - Note that only for $\lambda\geq \sqrt{\frac{10V}{a}}$
the cross section in the two directions, namely $\theta\sim 0$ and
$\theta$=$\frac{\pi}{2}$, $\varphi$=$\frac{\pi}{2}$ have the same order
and for all other $\lambda$ the expression~(\ref{dopaeo}) has always
advantages.\\
\hspace*{1cm}Finally, note that in Ref.~\cite{wkdelogi} using Feynman
perturbation
tehniques authors have analyzed the conversion of GWs
into EM waves in an uniform electrostatic and magnetic
fields in which their background are confirmed to the region between
the planes $z=-l/2$ and $z=l/2$. Authors have considered the only
case in which both momenta of graviton and photon are parallel to
the z axis ( this case corresponds with $\theta =0$ in our results).
Therefore, we see that our results are more realistic. Finally we
note again that the cross sections of EM- gravitational conversion
may have the observable value in the present technical scenario.\\
\hspace*{1cm}In this work we considered only a theoretical basis for
experiments.
Other problems with detection will be investigated in the future.\\

{\Large \bf Acknowledgements}\\
\hspace*{1cm} H. N. L would like to thank Prof. T. Yukawa and Theory
Group KEK, where this work was initiated for hospitality.
He also thanks the National Basic
Research Program on Natural Science for support under the contract
No. KT-04-3.1.12. T. A. T would like to thank Professor J. Tran
Thanh Van for partial support and he also would like to thank Professor
Abdus
Salam, UNESCO, and IAEA for hospitality at the International Centre for
Theoretical Physics, Trieste, Italy, where a part of work was done.
 D.V.S thanks Prof. N.S.Han for encouragement.\\

\end{document}